\documentstyle[aps,prc,preprint,tighten,epsf]{revtex}

\newcommand{\bea}{\begin{eqnarray}}
\newcommand{\eea}{\end{eqnarray}}
\newcommand{\bnn}{\begin{eqnarray*}}
\newcommand{\enn}{\end{eqnarray*}}
\newcommand{\be}{\begin{equation}}
\newcommand{\ee}{\end{equation}}

\newcommand{\bfa}[1]{\mbox{\boldmath${#1}$}}

\begin{document}

\draft

\title{Study of relativistic bound state wave functions\\
in quasielastic $(e,e^\prime p)$ reactions}

\author{S.\ Ulrych and H.\ M\"{u}ther}
\address{Institut f\"ur Theoretische Physik, Universit\"at T\"ubingen,
         D-72076 T\"ubingen, Germany}

\maketitle

\begin{abstract}
The unpolarized response functions of the quasielastic 
$^{16}O(e,e^\prime p)^{15}N$ reaction are calculated for three
different types of relativistic bound state wave functions.
The wave functions are obtained from relativistic Hartree, 
relativistic Hartree-Fock
and density dependent relativistic 
Hartree calculations that reproduce the experimental
rms charge radius of $^{16}$O.
The sensitivity of the unpolarized response functions
to the single particle structure of the different models is investigated
in the relativistic plane wave impulse approximation. 
Redistributions of the momentum dependence in the
longitudinal and transverse response function can be related
to the binding energy of the single particle states.
The interference responses $R_{LT}$ and $R_{TT}$ reveal a strong 
sensitivity to the small component  
of the relativistic bound state
wave function. 
\end{abstract}

\pacs{PACS number(s): 21.30.Fe, 21.60.Jz, 24.10.Jv, 25.30.Fj}

\section{Introduction}
\label{intro}
The exclusive $(e,e^\prime p)$ proton knockout reaction is 
a powerful tool for the investigation
of the single particle structure of complex nuclei 
\cite{Frul85,Boff93,Kell96}.
Quasiparticle properties, such as occupation probabilities, 
spectroscopic factors, binding energies and momentum distributions 
can be determined and compared to the results of
theoretical models, which have to include both, information
about the electromagnetic reaction mechanism and the wave function
of the bound nucleon.\\
The nuclear structure contributions can be seen e.g.~in the
reduced cross section. 
In the nonrelativistic plane wave impulse 
approximation, the reduced cross section
is proportional to the momentum distribution of a single bound proton inside
the nucleus. This proportionality is characterized by the
spectroscopic factor, which corresponds to the probability
that a single particle state is occupied by a nucleon. 
Therefore, spectroscopic factors can be extracted 
comparing experimental scattering data with theoretical predictions for
the momentum distribution of the nucleus.  
In a mean field calculation the spectroscopic factor is 
equal to one for occupied or zero for unoccupied states, respectively. 
Different spectroscopic factors indicate 
the deviation from the shell model picture and therefore
the importance of nucleon-nucleon correlations.\\ 
The theoretical background of the scattering formalism
is provided by quantum electrodynamics,
a complete relativistic framework, 
which describes the
electromagnetic interaction with highest accuracy.
Consequently, all contributions to the scattering
amplitude especially the hadronic current operator are relativistic
expressions and the matrix elements of the current 
should be calculated
between states obtained from a relativistic
treatment of the many-body problem.
Due to the higher complexity of the relativistic 
problem the solution of a nonrelativistic reduction 
is more economic and it is able to reproduce 
the experimental data in a wide range  
of missing energies and momenta.\\ 
Relativistic calculations of 
exclusive $(e,e^\prime p)$ scattering reactions, 
including electron distortion and
final state interactions, have been
performed \cite{Mcd90,Udia93,Jin92}. 
The results provided spectroscopic factors e.g. for
the $3s_{1/2}$ and $2d_{1/2}$ shells in $^{208}Pb$ of
$S_\alpha \simeq 0.7$, consistent with earlier theoretical
predictions.  
Though these spectroscopic factors are extracted 
from the low $p_m$ data $(p_m \leq 300 MeV)$, 
relativistic calculations can give simultaneously a good 
reproduction of the high $p_m$ data
\cite{Udia96}, where the main effects arise
from the improved relativistic treatment 
of the electron distortion and the final state interaction.\\
The inclusion of electron distortion and
final state interactions can give a good description
of the experimental data, but for a deeper understanding of
the relativistic reaction mechanisms the relativistic plane wave impuls
approximation (RPWIA) seems to be an appropriate calculation scheme.  
Recently, the RPWIA was chosen to study the role
of the negative energy components of the bound nucleon
wave functions \cite{Caba98}. Analyzing
the factorization of the scattering cross section, a feature
of the nonrelativistic limit, the importance
of relativistic effects for different choices of current operators, 
kinematics and restorations of current
conservation has been investigated thoroughly.
All calculations were made using the bound state
wave function of Horowitz and Serot \cite{Sero79,Horo81,Horo91}. Therefore it is
interesting to ask,   
what influence on the results can be observed, if bound
state wave functions from other relativistic nuclear
structure calculations are considered. This question shall be
studied in this paper.\\
For the investigation bound state wave functions from three different
approaches were chosen, namely, the Relativistic Hartree (RH) approach 
of Horowitz and Serot \cite{Sero79,Horo81,Horo91}, the
Relativistic Hartree-Fock calculations (RHF) of Bouyssy \textit{et al.}
\cite{Bouy87} and a Density Dependent Hartree (RDDH) approach
including rearrangement terms of Fuchs and Lenske \cite{Fuch95}.
Each of them was used to calculate the response functions
of the $^{16}O(e,e^\prime p)^{16}N$
electron scattering reaction.\\   
All models used are based on a microscopic understanding
of the nucleus using neutrons and protons as effective
degrees of freedom and the exchange of $\sigma$-, $\omega$-, $\rho$- and,
in the RHF approximation, $\pi$-mesons to mediate the nuclear force. 
They provide a consistent mathematical description 
starting from a covariant Lagrangian.
In these models the most important contributions 
to the nucleon-nucleon potential  
arise from an attractive $\sigma$-meson exchange, which is  
understood as a parametrization of the $2\pi$ exchange
diagrams, and a repulsive $\omega$-meson. 
Calculations in finite nuclei 
within the mean field approximation
are characterized by two large potentials of scalar and vector type
(S-V) cancelling each other to a large amount. 
As a result the spin orbit splitting emerges 
automatically.\\
The RH and the RHF approaches use phenomenological
one boson potentials.  
In both models the model parameters are determined
in essentially the same way. The $\sigma$-N and
$\omega$-N coupling constants and the $\sigma$-meson mass
were fixed in both cases to reproduce the saturation point in nuclear matter
and the charge rms radius of $^{16}O$.\\
The third model is the RDDH approximation, which is a first
step to the relativistic description of finite
nuclei using realistic forces.  
In the RDDH approximation
the relativistic Brueckner-Hartree-Fock (RBHF) potential
of a realistic interaction (G matrix) is parametrized in nuclear matter 
in terms of density
dependent coupling constants and the parametrized interaction
is then applied to finite systems.  
This calculation scheme can be regarded as a reliable
approximation for the self consistent solution
of the relativistic Brueckner-Hartree-Fock equations
in finite nuclei \cite{Frit93}. 
RDDH calculations were 
performed by Brockmann and Toki \cite{Broc92}. It 
was extended to the Hartree-Fock 
approximation by Fritz and M\"uther \cite{Frit94} and 
Boersma and Malfliet \cite{Boer94}.
In the work presented here, we use bound state wave functions from
the vector density dependent RDDH approach
of Ref. \cite{Fuch95}.  
This many body approximation accounts for additional
rearrangement terms arising from the field theoretical
argument that the density dependent coupling
constants have to be considered as functionals of the baryon fields.
Due to the rearrangement terms this model
goes beyond an effective
Brueckner Hartree-Fock approximation.
The model was chosen from other RDDH calculations
due to its good results for the single particle as well as
the bulk properties of $^{16}$O.\\  
The results for the three nuclear structure calculations
yield essentially identical results for the charge radii of $^{16}O$,
but have a different single particle structure.
The single particle structure can
be tested directly with $(e,e^\prime p)$ knockout reactions.
Using this reaction we want to investigate 
the sensitivity of the
unpolarized response functions in the RPWIA to the model functions
and to relativistic effects included in the bound state
wave functions of the three approaches.

\section{Electron scattering observables}
As stated in section \ref{intro}, the theoretical description 
has to account for the scattering formalism 
as well as for the description of finite nucleon systems.
In the following section the theory will be summarized, which is 
needed for the electron scattering observables 
calculated in the present work. 
The theoretical foundations for exclusive electron scattering 
reactions were
developed in several publications \cite{Don841}-\cite{Pick89}.
The present paper follows the conventions 
of \cite{Pick89} except to some changes
in the notation. 
The nuclear information is included  
in the matrix elements of
the
hadronic current 
\be
J^\mu(q)=\langle\, p_x s_x , \psi_{fP_B}\,\vert
\,\hat{J}^\mu(q)\,\vert\, \psi_{iP_A}\,\rangle\;.
\ee
The initial state $\psi_{iP_A}$ describes the $A$ particle
target nucleus with the four momentum
$P^\mu_A=(M_A,\bfa{0})$, whereas the final state consists of the $A-1$ particle
residual nucleus $\psi_{fP_B}$ 
$(P^\mu_B=(E_B,\bfa{p}_B))$ and the ejected proton
$(p^\mu_x=(E_x,\bfa{p}_x))$. 
The present investigation will be performed in the RPWIA, i.e.,
the outgoing proton is represented by a 
plane wave $\overline{u}(\bfa{p}_x,s_x)$
and the current is given by the 
current of an individual constituent, treated as 
a free particle. The full complexity of the initial
and outgoing states for the bound nucleon system
is therefore replaced by a single shell model
wave function $\psi_\alpha(\bfa{p})$. In this approximation one can write 
for the above 
current matrix elements 
\be
J^\mu(q)=\overline{u}(\bfa{p}_x,s_x)\hat{J}^\mu(q)\psi_\alpha(\bfa{p})
\;.
\ee 
The calculation is done completely in momentum space. 
The bound state wave function
$\psi_\alpha(\bfa{p})$ for the initial state is the 
information we obtain from the shell model calculations 
and $\bfa{p}=\bfa{p}_m$ corresponds to the missing momentum.
The results of the approximation, neglecting proton and
electron distortion, should not be compared to the experimental data.
This approximation, however, should allow us to study the
sensitivity of the results on the model for the
bound state wave function $\psi_\alpha(\bfa{p})$.\\ 
In the present investigation the main attention
will be paid to 
the response functions, which are directly related
to the hadronic tensor. 
If polarization
is not taken into account
for the incoming electron beam,
as well as for the target and the outgoing particles, 
we can define the hadronic tensor including an 
average over the initial states and a sum over the
final states 
\begin{equation}
\label{tensor}
W^{\mu\nu}=\overline{\sum_i}\, \sum_f\hspace{-0.5cm}\int\;\,
              \delta(\epsilon_f-\epsilon_i-\omega)\; J^{*\mu}(q)\,J^\nu(q)\;.
\end{equation}
In this definition an integration over the momentum
conserving delta function has 
already been performed, which provides the relation
$\bfa{q}=\bfa{p}_x-\bfa{p}$.
The response functions can be constructed using the
hadronic tensor according to
\begin{eqnarray}
\label{response}
R_L&=&\int_{line} d\epsilon_{p_x}\, W^{00}\;, \nonumber\\
R_T&=&\int_{line} d\epsilon_{p_x}\,( W^{11}+W^{22})\;,\\
R_{LT}\cos{\phi}&=&\int_{line} d\epsilon_{p_x}\,(-W^{01}-W^{
10})\;,\nonumber\\
R_{TT}
\cos{2\phi}&=&\int_{line} d\epsilon_{p_x}\,(W^{11}-W^{22})\;,\nonumber
\end{eqnarray}
including an integration over a linewidth in the missing
energy spectrum. 
These response functions can be used to express the
unpolarized scattering cross section as a contraction of
hadronic responses and the appropriate electron 
contributions, which are defined as in \cite{Kell96,Pick87} 
\be
\frac{d\sigma}{d\epsilon_{k^\prime}d\Omega_{k^\prime}d\Omega_{p_x}}
=\frac{m\vert \bfa{p}_x \vert}{(2\pi)^3}\;\sigma_{M}
\left(V_LR_L+V_TR_T
+V_{LT}R_{LT}\cos{\phi}+V_{TT}R_{TT}
\cos{2\phi}\right)\;.
\ee
Since recoil effects of the residual nucleus
will not be considered in the present work,
an appropriate factor has been neglected in the above formula. 
Using the definitions of this section the observables 
depend, except to the bound state wave function,
on standard expressions
like the current operator or the Dirac spinor.

\section{Relativistic bound state wave functions}
It was shown in the last section that  
the information for the calculation of the
current matrix elements, which is 
required from relativistic nuclear structure
calculations, is the momentum space wave function $\Psi_{\alpha}(\bfa{p})$
of a particular shell model state $\alpha$.
The solutions of a relativistic Hartree or Hartree-Fock
calculation in finite nuclei are usually given in coordinate space.
The corresponding momentum space wave function can be obtained
by a Fourier
transformation according to
\be
\Psi_{\alpha}(\bfa{p})=\frac{1}{(2\pi)^{3/2}}
\int\!d^3\bfa{r}\;e^{-i\bfa{p}\cdot\bfa{r}}
\Psi_{\alpha}(\bfa{r})\;.
\ee
More details about the bound state wave functions are
given in Appendix A.
In the context of $(e,e^\prime p)$ scattering calculations  
a nuclear structure model should provide
a good reproduction of the charge distribution,
which can be observed directly in electron scattering experiments
at lower energies. The three nuclear structure models chosen
for the investigation here satisfy
this boundary condition. We want to study the influence of the different single
particle structure of these models
on the unpolarized response functions of Eq. (\ref{response}).\\ 
The models we use for the study are the relativistic Hartree approximation
(RH) of Ref.~\cite{Horo81}, the
relativistic Hartree-Fock (RHF) approximation of Ref.~\cite{Bouy87}
and the density dependent Hartree (RDDH) approximation of Ref.~\cite{Fuch95}.\\ 
The RH and the RHF models follow the same way to fix their free 
parameters. The $\sigma$-N and
$\omega$-N coupling constants and the $\sigma$-meson mass
are adjusted to reproduce both, the saturation point of nuclear
matter and the charge rms radius of $^{16}O$.\\ 
In the Hartree-Fock approximation 
the Dirac structure of the nucleon self energy $\Sigma(r)$
is modified due to the nonlocal structure of the two particle
interaction. 
The self energy $\Sigma(r)$, which incorporates the
influence of all other nucleons in the nucleus on
a single particle, is included in the free Dirac equation 
according to 
\be
(i\partial_\mu\gamma^\mu -m -\Sigma (r))\Psi_\alpha(\bfa{r})=0\;.
\ee
In a nuclear system, which is characterised by rotational 
invariance, parity conservation
and time reversal invariance,
the general structure of self energy is given by
\be
\label{structure}
\Sigma (r)=\Sigma_s (r)-\gamma_0\Sigma_0 (r) + \bfa{\gamma}\cdot \bfa{\Sigma}_v
(r)\;. 
\ee
In the Hartree approximation the spatial  
part of the vector self energy $\bfa{\Sigma}_v$ vanishes.
The remaining structure is normally 
considered as a potential of scalar and vector (S-V) type.
In the Hartree-Fock approximation
the additional $\bfa{\Sigma}_v$ term arises and gives 
a structural modification
compared to the mean field approximation.\\ 
In the Hartree-Fock calculations the potentials are
not given in the separated form of Eq. (\ref{structure}). 
In principle, the Dirac structure
can be projected out if the Fock contributions are redefined
in terms of local single particle potentials \cite{Bouy87} and incorporated
in the Dirac equation. 
From the nuclear matter calculations we know,
that in the Hartree-Fock approximation, every meson contributes to
$\Sigma_s$, $\Sigma_0$ and $\bfa{\Sigma}_v$, and that the
$\bfa{\Sigma}_v$ term is small. This should remain true also
in finite systems.\\ 
The models for the Hartree and the Hartree-Fock
approximation are described by the following parameter sets:
The $\sigma$-meson mass, especially sensitive to the rms radius,
is chosen as $m_\sigma=520$ MeV for the RH model \cite{Horo81}
and $m_\sigma=440$ MeV for the RHF model \cite{Bouy87}. The coupling constants
$g_\sigma$ and $g_\omega$
were fixed to reproduce the saturation point in nuclear matter.
In the RH as well as the RHF model the 
masses of the $\omega$- and $\rho$-meson are 
$m_\omega=783$ MeV and $m_\rho=770$ MeV in accordance
with the experimental values.
In the RH model \cite{Horo81} the $\rho$-meson coupling constant  
$g^2_\rho/{4\pi}=5.19$ is fixed to yield a bulk symmetry energy of $35$ MeV. 
For the Hartree-Fock calculation we have chosen the parameter set
$(e)$ of Ref.~\cite{Bouy87}, which provides the best results for
the bulk and single particle properties of $^{16}O$.
In this set the $\pi$-N and $\rho$-N coupling constants of the
Hartree-Fock calculations were fixed to the physical values
$f^2_\pi/{4\pi}=0.08$ and $g^2_\rho/{4\pi}=0.55$, where a
ratio of $f_\rho/g_\rho=3.7$ for the tensor
coupling of the $\rho$-meson was chosen.
The $\pi$-meson mass in this model is $m_\pi=138$ MeV.\\  
For the effective relativistic Brueckner-Hartree-Fock calculation
in finite nuclei we have chosen the
vector density dependent (VDD) Hartree approach
of Ref.~\cite{Fuch95}. 
Here, for the $\omega$-meson and 
$\rho$-meson mass as well as for the $\rho$-meson
coupling constant, the values of \cite{Horo81} were
adopted. For the $\sigma$-meson $m_\sigma=550$ MeV
was used. The coupling constants $g_{\sigma}$ and $g_{\omega}$
were taken from a parametrization of the 
nuclear matter DBHF self energies \cite{Broc90} with 
second order polynomials \cite{Hadd93}.\\
The single particle and bulk properties of the three models, which
provide the bound state wave functions needed in the electron
scattering calculations, 
are summarized in Tab.~\ref{tab:tab1}.  
The calculated energy per nucleon has been corrected in the RH
and RHF models to
account for the effects of the spurious center of mass motion.
A contribution of $0.67$ MeV has been added to the binding energy. 
In the RDDH model the results of Ref.~\cite{Fuch95}
include already the center of mass motion. 
Since the calculations for the RH and
the RHF approach
have been performed with the computer code developed by
Fritz and M\"uther \cite{Frit94}, 
the results of Tab.~\ref{tab:tab1} 
differ slightly from the original values given
in the Refs. \cite{Horo81,Bouy87} due to numerical
uncertainities. 
In the calculation scheme of Ref.~\cite{Frit94} 
the Hartree or Hartree-Fock Hamiltonian 
is diagonalized in a box basis.

\section{Bound states and response functions}
\begin{figure}
\epsfysize=9.0cm
\begin{center}
\makebox[6.6cm][c]{\epsfbox{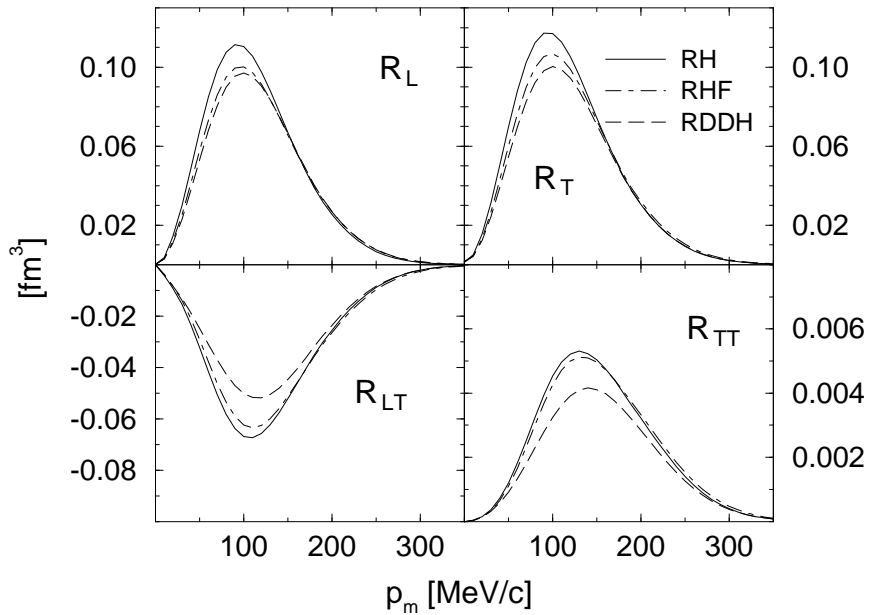}}
\end{center}
\caption{Response functions $R_L$, $R_T$,
$R_{LT}$ and $R_{TT}$ of Eq.~(\ref{response}) (in $[fm^3]$) 
as a function of the missing momentum
$p_m$ $[MeV/c]$. The solid line
corresponds to the RH, 
the dash-dotted line to the RHF and the
dashed line to the RDDH approach.}
\label{fig:one}
\end{figure}

\begin{figure}
\epsfysize=9.0cm
\begin{center}
\makebox[6.6cm][c]{\epsfbox{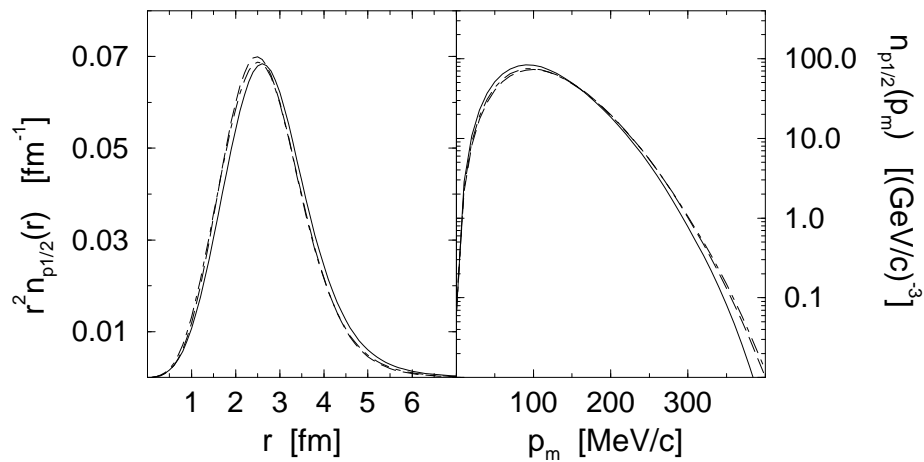}}
\end{center}
\caption{Left panel: density distribution $r^2n_{p1/2}(r)$ in $[fm^{-1}]$
as a function of the radius $r$ $([fm])$. Right panel: density
distribution in momentum space $n_{p1/2}(p_m)$ in $[GeV/c^{-3}]$ 
as a function of the missing momentum.
The lines are denoted as in Fig.~\ref{fig:one}.}
\label{fig:two}
\end{figure}

\begin{figure}
\epsfysize=9.0cm
\begin{center}
\makebox[6.6cm][c]{\epsfbox{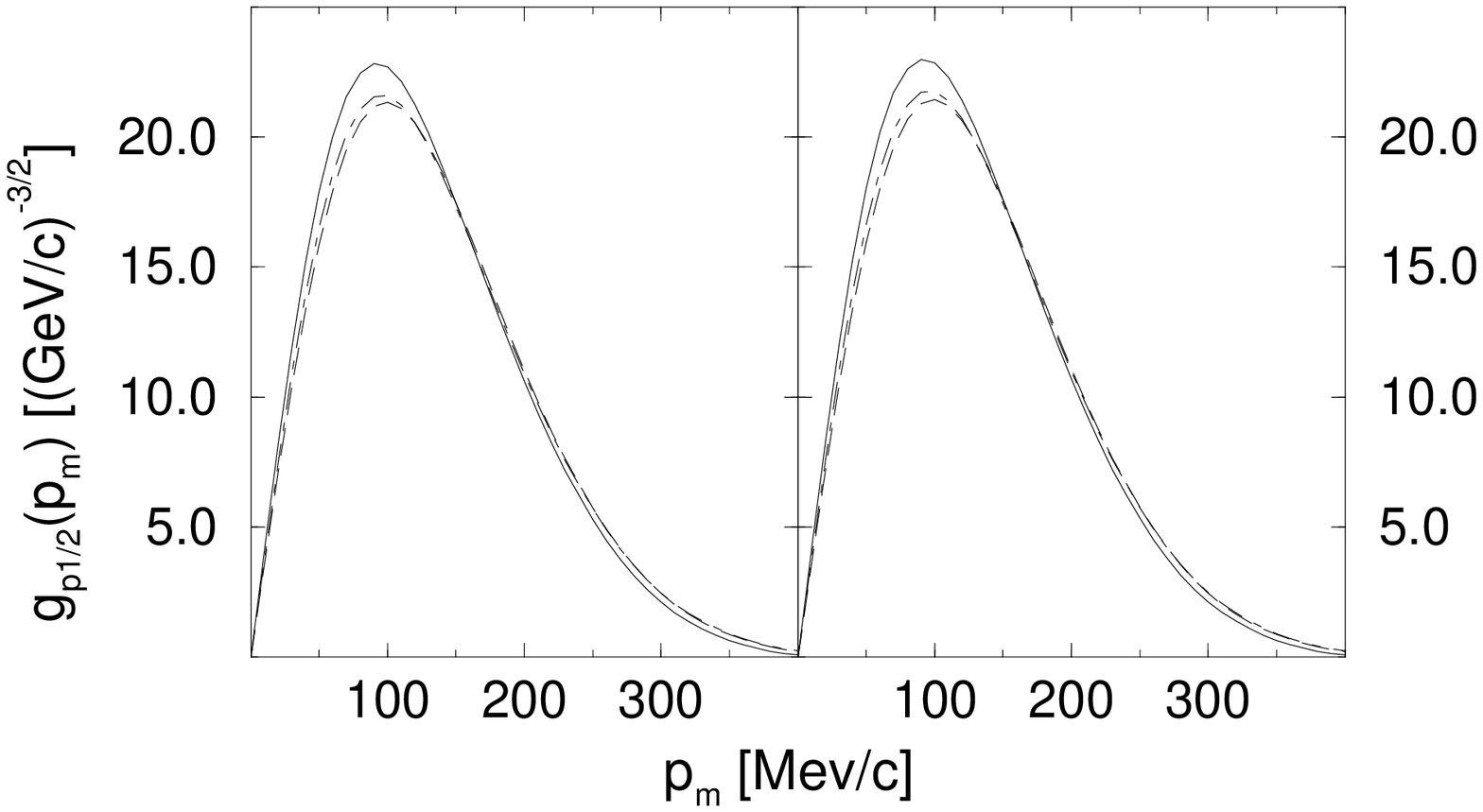}}
\end{center}
\caption{Left Panel: 
upper component $g_{p1/2}(p_m)$ of the relativistic bound state 
wave function in [$(GeV/c)^{-3/2}]$ 
as a function of the missing momentum. 
Right panel: nonrelativistic limit 
$g_{p1/2}^{nr}(p)$. The lines are denoted as in Fig.~\ref{fig:one}.} 
\label{fig:three}
\end{figure}

\begin{figure}
\epsfysize=9.0cm
\begin{center}
\makebox[6.6cm][c]{\epsfbox{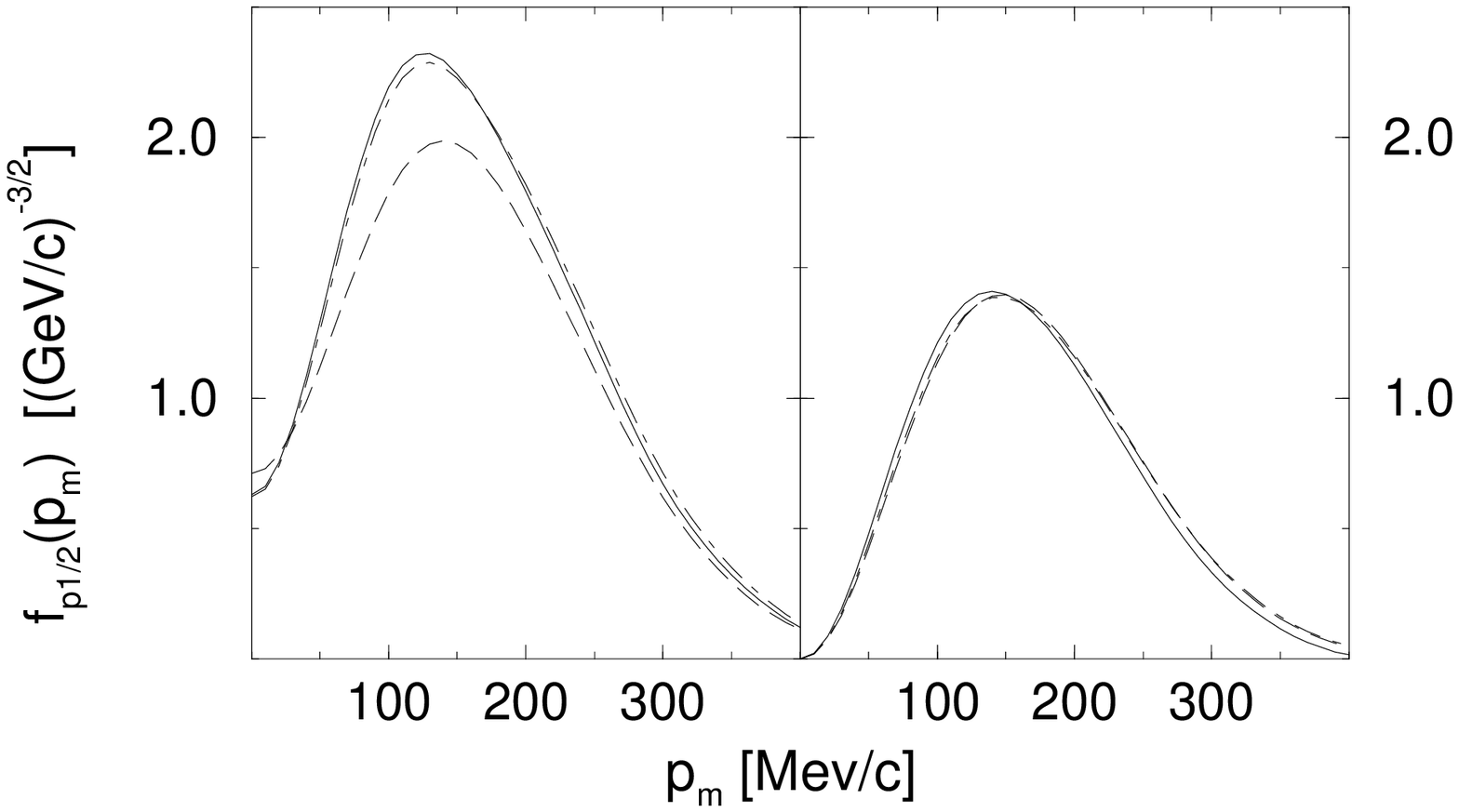}}
\end{center}
\caption{Left panel: lower component $f_{p1/2}(p_m)$ 
of the relativistic bound state
wave function in [$(GeV/c)^{-3/2}]$
as a function of the missing momentum. 
Right panel: nonrelativistic limit
$f_{p1/2}^{nr}(p)$. The lines are denoted as in Fig.~\ref{fig:one}.}
\label{fig:four}
\end{figure}

\begin{figure}
\epsfysize=9.0cm
\begin{center}
\makebox[6.6cm][c]{\epsfbox{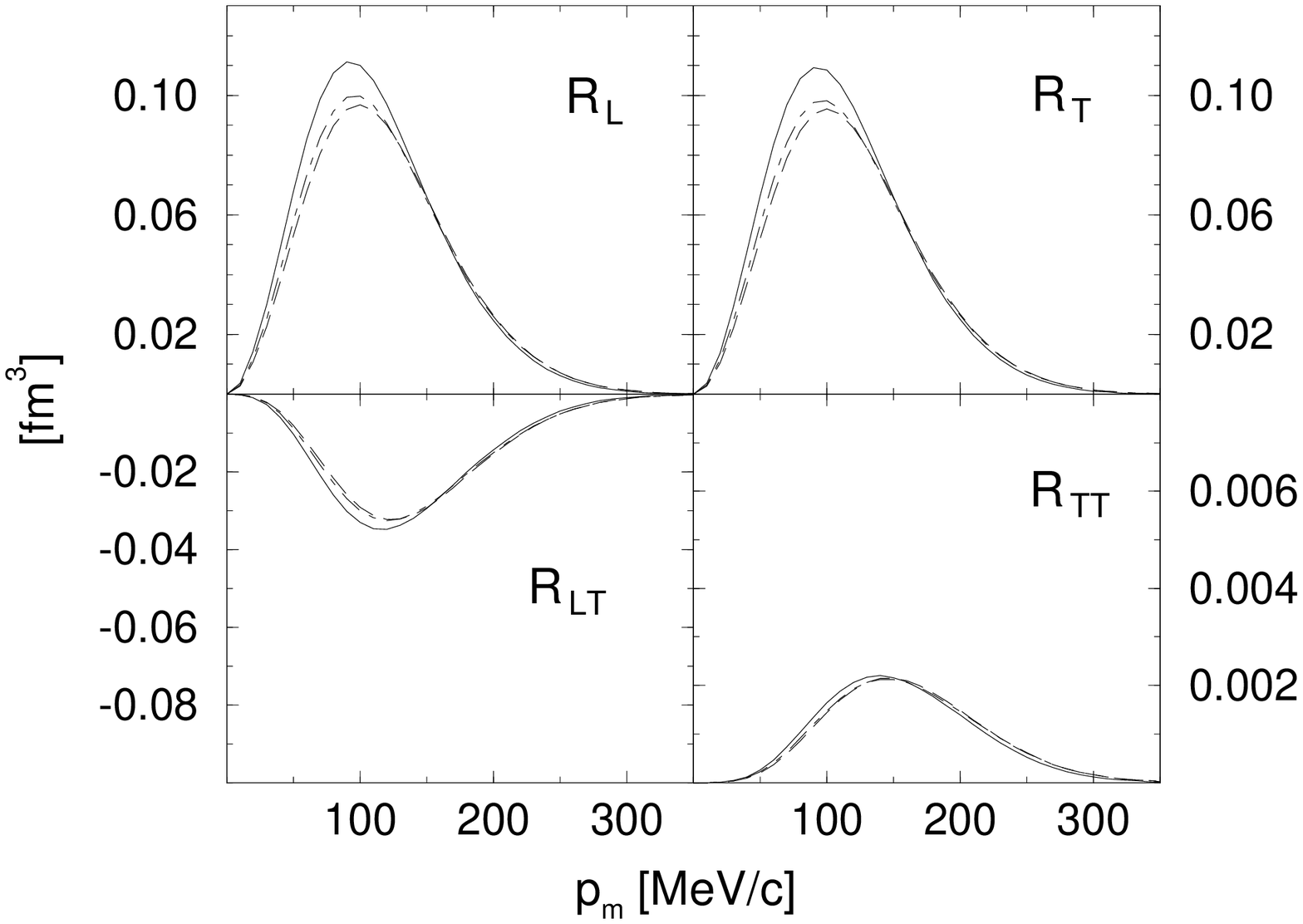}}
\end{center}
\caption{Response functions in the nonrelativistic limit. 
The labelling is the same as in Fig. \ref{fig:one}.}
\label{fig:six}
\end{figure}

The reaction considered in the present investigation
is the proton knockout reaction
$^{16}O(e,e^\prime p)^{15}N$, where the residual nucleus
is left in its ground state, i.e.~the proton is
knocked out of the $p_{1/2}$ state of $^{16}O$.\\
To calculate the response functions of the $(e,e^\prime p)$ reaction
the bound state wave functions of the three models discussed above 
were used. 
The computations have been performed 
completely in momentum space. A new computer code
has been developed.  
Note, that for the definition of our interference response 
functions in Eq. (\ref{response}) we use a different convention.
For the calculation of the knockout reaction
we have chosen
perpendicular kinematics, which provides us with the
interference terms of the relativistic response functions
$R_{LT}$ and $R_{TT}$. For the current operator
we take the CC1 description \cite{deFo83}, using the common
dipol form factors \cite{Kell96}. 
It was shown that this operator gives 
a simultaneous good description for 
high as well as low missing momenta \cite{Udia96} in $^{208}Pb$.  
Furthermore, this operator
enhances the contributions from the negative energy projections
of the bound state wave functions as stated in Ref.~\cite{Caba98}. 
For the four momentum transfer $q^\mu=k^\mu-k^{\prime\mu}=(\omega,\bfa{q})$  
we take kinematics I of Ref. \cite{Caba98} with
\be
\vert \bfa{q}\vert=500\; \mathrm{MeV/c} \hspace{2cm} \omega=131.56 
\;\mathrm{MeV}\;, 
\ee
where the value of $\omega$ corrsponds to the quasielastic
peak value $\omega_{QE}$.
Note, that for the definition of our interference response
functions in Eq. (\ref{response}) we use a different convention.\\
The response functions for the three models presented in
the last section are displayed in Fig.~\ref{fig:one}. 
On the upper left panel we see the longitudinal response
function $R_L$. 
In all figures the results obtained 
from RH bound state wave functions are displayed with solid lines,
whereas the RHF and the RDDH results are displayed
with dash-dotted and dashed lines, respectively.
It can be observed that the maximal strength of $R_L$ 
is reduced by the RHF and the RDDH approach in the order of  
$10\%$ relative to the RH approach.\\
The integrated strength of the response
functions 
\be
\label{intresponse}
I_i=\int dp_m p^2_m R_i(p_m) \;,
\ee
($i=L,T,LT,TT$) shows if either a reduction corresponds
to a redistribution of the strength in momentum space or there
occurs a true reduction of the strength.
The results for the integrations are displayed in
Tab.~\ref{tab:tab2}.
For the longitudinal 
response function one finds values of $I_L=2.81$ for the RDDH and
$I_L=2.83$ in the case of the RH and the RHF approach, where
the factor $100$ has been multiplied to the original values.
Therefore, the reduction of the
maxima can be understood essentially as a redistribution of the
total strength over the momentum scale, leading to higher
values in the RHF and RDDH response functions for large $p_m$.\\
For the transversal response $R_T$ displayed in the upper right panel
of Fig.~\ref{fig:one} 
a similar behaviour as for $R_L$ is found considering the
RH and the RHF bound state wave functions.
However, the response function $R_T$ obtained from
the RDDH bound state wave functions shows 
a stronger reduction relative to
the other transversal responses. This becomes more obvious, 
if we consider the integrated strength, which takes $I_T=3.30$
for the RH and $I_T=3.31$ for the RHF response, whereas
for the RDDH approach we find $I_T=3.12$.\\ 
In the interference contributions
this integral reduction can be observed as well. In addition a
small reduction is induced by the RH wave functions compared
to the response $R_{LT}$ of the RHF wave functions,
which is reflected by a smaller
value of $I_{LT}$. This reduction is even enhanced in the
$R_{TT}$ response. 
Note, that the maximum values arising from the RH wave
functions are larger. That means we have an additional enhancement
of the large momentum components in the RHF transverse responses
arising from the increased total strength,
which is not so obvious in Fig.~\ref{fig:one}.\\
In order to summarize these features, one may say that we
observe two effects
\begin{itemize}
\item a redistribution of the strength of the response functions
from lower momenta to high missing momenta if we compare results
derived from RH, RHF and RDDH, respectively.
\item a reduction of the integrated strength if we compare the 
results of RH and RHF on one side with these derived from RDDH
on the other side. This reduction is negligible for the
longitudinal response $R_L$, but significant for the 
response functions $R_{LT}$ and $R_{TT}$. 
\end{itemize}
In order to understand the redistribution of the
strength in the example of the longitudinal
response $R_L$, we consider 
the single particle densities of the three models in
the momentum and the coordinate space.
The single particle density is defined in momentum space
according to
\bea
\label{dens}
n_\alpha(p)&=&\frac{2j_\alpha+1}{4\pi}\int\! d\Omega_p\, 
\overline{\Psi}_\alpha(\bfa{p})\gamma_0\Psi_\alpha(\bfa{p})\\ 
&=&\frac{2j_\alpha+1}{4\pi}\left(
g_{\alpha}^2(p)+ f_{\alpha}^2 (p)\right)\;,\nonumber
\eea
where the upper and lower components 
of the relativistic wave functions, $g_{\alpha}(p)$
and $f_{\alpha}(p)$, are defined in Appendix A. 
The density in coordinate space is defined in analogy to
Eq.~(\ref{dens}) with the
Fourier transformed expressions.
The densities are normalized to the number of particles 
in a particular shell $\alpha$.\\
On the left panel of Fig.~\ref{fig:two} the density in coordinate space 
multiplied with 
the square of the radius, $r^2n_{p1/2}(r)$, is shown. 
For this expression the area below the curves is identical for
the three different models. The density $n_{p1/2}(r)$ itself  
shows an analog but inverse behaviour as the longitudinal
response function $R_L$. The function $r^2n_{p1/2}(r)$ indicates
that the single particle 
densities of the RHF and the RDDH approaches
are shifted to smaller radii. This is the reason of the higher
momentum components in the wave functions. Though the shift 
in the radial wave functions is small,
we have seen that this leads to a $10 \%$ reduction of the maxima
of $R_L$. 
This redistribution of the momentum distribution and density distribution
in configuration space is also connected to the single particle
energy of the $p1/2$ state (see table \ref{tab:tab1}). This state
is most weakly bound in the RH approximation. Therefore it is
less localized in configuration space and the momentum distribution
of this state is shifted to smaller momenta in this approach
as compared to the other calculations.\\ 
In the nonrelativistic PWIA the momentum density of a 
single particle state  
is proportional to the reduced cross section.
On the right panel of Fig.~\ref{fig:two} the momentum density $n_{p1/2}(p_m)$
is shown in a logarithmic scale, which emphasizes the high momentum 
contributions. Corresponding to the
longitudinal response function the maxima of $n_{p1/2}(p_m)$ 
are reduced starting from $n_{p1/2}^{max}=83.3$ for the
RH wave functions to $n_{p1/2}^{max}=74.8$
respectively $n_{p1/2}^{max}=73.0$  
for the RHF and the RDDH wave functions (all values in $[(GeV/c)^-3]$).
The numerical values show that the behaviour of the 
momentum density
explains the structure of the response function $R_L$. 
In Fig.~\ref{fig:three} we can see
in addition that the
upper component of the bound state wave function $g_{p1/2}(p_m)$
already reflects the typical momentum distribution of $R_L$.\\
In the RH and the RHF responses
beside the effects arising from the different single 
particle densities no significant differences 
can be observed. 
The $\bfa{\Sigma}_v$ term, included in the RHF potential, 
can therefore be expected to give no substantial modification 
of the response functions.\\
The reduction of the integrated RDDH responses $R_T$, $R_{LT}$ and $R_{TT}$
has another origin. This reduction is closely related to
the lower component of the bound state wave function $f_{\alpha}$.
In relativistic nuclear structure calculations this small
component is enhanced as compared to a small component which
just arises from a boost of a nucleon with mass $m$. In order
to explore the sensitivity we compare results of the relativistic
calculation, with those in which this enhancement of the small
component is suppressed.
This nonrelativistic limit $\Psi_{\alpha}^{nr}(\bfa{p})$ of the bound
state wave function is
defined as
\be
\label{nrel}
g_{\alpha}^{nr}(p)=N^{nr}g_{\alpha}(p)\hspace{2cm}
f_{\alpha}^{nr}(p)=N^{nr}\frac{p}{E_p+m}\,g_{\alpha}(p)\;,
\ee
where $N^{nr}$ is a normalisation constant and $E_p=\sqrt{p^2+m^2}$. 
The small component $f_{\alpha}^{nr}(p)$ is just a result 
of boosting the single particle spinor and therefore of
pure kinematical origin. 
This definition (\ref{nrel}) means that 
the momentum dependence of the lower component $f_{\alpha}^{nr}(p)$
is given, except to a relativistic factor,
by the upper component $g_{\alpha}(p)$ of the relativistic 
wave function.
On the right panel of Fig.~\ref{fig:three} the upper component
of this nonrelativistic limit is shown. 
The relativistic upper component $g_{p1/2}(p)$ is slightly smaller than 
than the nonrelativistic $g_{p1/2}^{nr}(p)$ due to the
normalisation constants $N^{nr}$.
The situation is completely different for the lower
component $f_{p1/2}(p)$, which is displayed on the left
panel of Fig.~\ref{fig:four}.
We find an enhancement of approximately $60\%$ compared to
the nonrelativistic wave function $f_{p1/2}^{nr}(p)$ shown on
the right panel of Fig.~\ref{fig:four}. 
This enhancement of the lower component for relativistic nucleons
in the medium is consistent with the results found in
nuclear matter, where the enhancement of the lower
component of the Dirac spinor is characterized by an effective mass of
approximately $m^*=m+U_S=600$ MeV compared to $m=938.9$ MeV for free
nucleons. This additional medium dependence is typical
for relativistic calculations and it is not included
in a nonrelativistic reduction.\\ 
For the relativistic lower component $f_{p1/2}(p)$,
the bound state wave function 
of the RDDH approach differs significantly from
the lower components of the two other approaches.
This indicates that the scalar potential $U_S$ of the RDDH
calculations is less attractive.  
We have seen that in the interference response
functions $R_{LT}$ and $R_{TT}$, shown in Fig.~\ref{fig:one}, 
these differences modify the results notably, whereas
the transversal response function $R_T$ shows small
changes, since it is dominated by the upper component
of the bound state wave function.\\ 
The sensitivity of the interference responses 
on relativistic effects in the bound state wave functions
can also be demonstrated if the response functions are calculated
in the nonrelativistic limit.
This is shown in Fig.~\ref{fig:six}. A strong reduction 
especially in the reponses $R_{LT}$ and $R_{TT}$ can be
observed. The strong dependence on the lower component has
already been shown in Ref.~\cite{Caba98}.
In addition, we can see here that 
the model dependent information, included
in the lower component, disappears. 
The differences of the RDDH responses compared to the
other approaches are strongly reduced.
The same effect as discussed above can be observed in the
nonrelativistic limit of the RDDH lower 
component $f_{p1/2}^{nr}(p)$ displayed in Fig.~\ref{fig:four}.
In this case the lower components now simply reflect 
the behaviour of the upper components.\\ 
In order to learn more about the typical structure
of bound state wave functions, 
the calculations
have also been performed with the density dependent
models of Fritz and M\"uther \cite{Frit94}. This calculation scheme provides
a good description of the binding energies, whereas the
charge radii are too small. The wave functions are therefore
inadequate for a study in the context of
$(e,e^\prime p)$ reactions and the results are not
presented in detail. Nevertheless, two additional
informations can be obtained from these calculations. 
The same structural difference of the longitudinal response $R_L$
has been found between the Hartree and the Hartree-Fock approximation. 
A reduction of the maxima of approximately $10\%$ and 
larger high momentum components.
This can also be traced back
to a different single particle structure. 
Furthermore, no reduction in 
the lower component of the wave function
can been observed, i.e., the reduction
in the RDDH approach of Ref. \cite{Fuch95}
is due to the rearrangement terms and
not part of the density dependent approach.

\section{Summary and Conclusions}
Bound state wave functions obtained from a relativistic
Hartree, Hartree-Fock and a density dependent Hartree approach 
were compared and the influence on the unpolarized
response functions of the $^{16}O(e,e^\prime p)^{15}N$ reaction was studied. 
The nuclear structure models chosen 
reproduce the experimental charge radius of $^{16}O$.
It was therefore possible to study the sensitivity
of the response functions to the different single
particle structure provided by the models.
The aim of this investigation is to give an estimate of possible
uncertainities in the response functions and to show
the influence of the additional information provided by
a relativistic treatment of the many body problem.\\ 
The calculations were performed in the relativistic
plane wave impulse approximation, where modifications
of the results can clearly be assigned to the bound
state wave functions.
Comparing the three models the
maxima of the $R_L$ and $R_T$ responses 
differ by $10\%-15\%$.
This reduction of the RHF and the RDDH response functions
is accompanied by an enhancement at larger missing
momenta, so that the integrated strength remains unchanged.
This behaviour can be connected with the density distribution
of the single particle states reflecting the single particle
binding energies.
The vector self energy $\bfa{\Sigma}_v$,
which appears in the Hartree-Fock approximation, has
no significant influence on the structure of the response functions.
This result was expected, since in nuclear
matter calculations its contributions are found to be small.\\ 
The interference responses $R_{LT}$ and $R_{TT}$  
are known to be sensitive to the lower component
of the relativistic bound state wave function.
The RDDH lower component is significantly reduced
compared to the wave functions of the other models.
This reduction is clearly measured by the interference
responses and influences also the integrated strengths
of the response functions. The maxima of $R_{LT}$ and $R_{TT}$
even show a variation of around $20\%$. In the nonrelativistic
limit this model dependent information is lost
and only small differences for the interference
responses can be observed. Especially, the interference
responses are therefore sensitive to a complete relativistic
description of the nuclear structure problem and can
be used to test different relativistic approximation schemes
of nuclear many body systems. 
\section*{Acknowledgments}
We would like to thank J.M.~Ud\'\i as very much for helpful and
interesting discussions. We further thank C. Fuchs for providing
us with the bound state wave functions of Ref.~\cite{Fuch95}.
This work has been supported by the Sonderforschungsbereich SFB 382
(DFG, Germany) and the 
``Graduiertenkolleg Struktur und
Wechselwirkung von Hadronen und Kernen'' (DFG GRK 132/2).
This support is gratefully acknowledged.

\section*{Appendix A}
\label{Bound}
This section gives a survey of the explicit definitions
for the relativistic bound state wave functions used
in the present work.
The solutions of the relativistic Hartree 
or Hartree-Fock equations are calculated in coordinate space and have the
general structure 
\be
\Psi_{\alpha}(\bfa{r})=
        \left(\begin{array}{@{\hspace{0pt}}c@{\hspace{0pt}}}
                g_{nlj}^{m_\tau}(r)Y_{lj}^{m_j}(\Omega_r) \\
                \; i f_{nl^\prime j}^{m_\tau} (r)Y_{l^\prime j}^{m_j}
                (\Omega_r)\;
                \end{array}\right)\chi^{m_\tau}\;.
\ee
$\alpha=\{n(ls)jm_j\tau m_\tau\}$ is an abbreviation for the 
various quantum numbers,
where $n$ denotes the radial quantum number, 
$l$ the angular momentum, $s$ the spin, $j$ the coupled
angular momentum, $\tau$ the isospin, $m_j$ and $m_\tau$ 
the orientation of the coupled angular momentum 
and the isopin.
The angular momentum of the lower component $l^\prime$ 
is given by the relation
\be
l^\prime=\left\{\begin{array}{@{\hspace{0pt}}c@{\hspace{0pt}}}
              \;  l+1 \,\,\,\, \mathrm{for} \,\,\,\, l=j-1/2 \:\\
              \;   l-1 \,\,\,\, \mathrm{for} \,\,\,\,l=j+1/2\; 
                \end{array}\right\}
\ee
The normalization of the states is 
\be
\label{norm1}
\int\!d^3\bfa{r}\,\Psi_{\alpha}^\dagger(\bfa{r})
\Psi_{\alpha}(\bfa{r})=\int_0^{\infty}
r^2 dr\;\left(
g_{\alpha}^2(r)
                + f_{\alpha}^2 (r)\right)=
1\;.
\ee
The Fourier transformation of the relativistic bound nucleon wave
functions is given by
\be
\Psi_{\alpha}(\bfa{p})=\frac{1}{(2\pi)^{3/2}} 
\int\!d^3\bfa{r}\;e^{-i\bfa{p}\cdot\bfa{r}}
\Psi_{\alpha}(\bfa{r})\; 
\ee
which leads to the following  
explicit form of 
the bound state wave function in momentum space 
\be
\label{boundstate}
\Psi_{\alpha}(\bfa{p})=
        \left(\begin{array}{@{\hspace{0pt}}c@{\hspace{0pt}}}
               (-i)^l\, g_{nlj}^{m_\tau}(p)\,Y_{lj}^{m_j}(\Omega_p) \\
                \; i(-i)^{l^\prime}\,
                 f_{nl^\prime j}^{m_\tau} (p)\,Y_{l^\prime j}^{m_j}
                (\Omega_p)\;
                \end{array}\right)\chi^{m_\tau}\;,
\ee
where the real amplitudes $g_\alpha(p)$ and $f_\alpha(p)$ are
defined according to
\bea
g_{nlj}^{m_\tau}(p)&=&\sqrt{\frac{2}{\pi}}\;\int_0^{\infty} 
r^2 dr\; j_l(pr)\, g_{nlj}^{m_\tau}(r)\nonumber\\
f_{nl^\prime j}^{m_\tau}(p)&=&\sqrt{\frac{2}{\pi}}\;\int_0^{\infty}
r^2 dr\; j_{l^\prime}(pr)\, f_{nl^\prime j}^{m_\tau}(r)\;       
\eea
and for the normalisation of the states
it follows that
\be
\int\!d^3\bfa{p}\,\Psi_{\alpha}^\dagger(\bfa{p})
\Psi_{\alpha}(\bfa{p})=\int_0^{\infty}
p^2 dp\;
\left(
g_{\alpha}^2(p)
+ f_{\alpha}^2 (p)\right)=
1\;.
\ee 
These momentum wave functions served as the input for 
the relativistic quasielastic electron scattering 
calculations. All calculations are performed in momentum space.
Therefore, the momentum space wave functions were multiplied
directly with the current operator and the Dirac spinor
to obtain the matrix elements of the hadronic current.

\vfil\eject
\begin{table}
\begin{tabular}{|c c c c c|}
\multicolumn{5}{|c|}{$^{16}$O}\\
  & RH  & RHF & RDDH & Exp.\\
&&&&\\
\hline
&&&&\\
$1s_{1/2}\;[MeV]$ & -37.2 & -38.6 & -36.0 & -40$\pm$8 \\
$1p_{3/2}\;[MeV]$ & -16.7 & -18.0 & -17.1 & -18.4 \\
$1p_{1/2}\;[MeV]$ &  -8.8 & -10.6 & -12.9 & -12.1 \\
$E/A\;[MeV]$      & -5.56 & -6.10 & -7.82 & -7.98 \\
$R_{ch}\;[fm]$   & 2.75 & 2.74 & 2.75 & 2.73 \\
&&&&\\
\end{tabular}
\caption{The proton single particle energies [MeV] for the
$1s_{1/2}$, $1p_{3/2}$ and $1p_{1/2}$ shells, the energy per nucleon
(E/A) [MeV] and the rms charge radius $R_{Ch}$ [fm] of $^{16}$O
for the relativistic Hartree (RH), the relativistic
Hartree-Fock (RHF) and the relativistic density
dependent Hartree (RDDH) approach.
The calculated energy has been corrected to
account for the effects of the spurious center of mass motion.
}
\label{tab:tab1}
\end{table}

\begin{table}
\begin{tabular}{|c c c c |}
\multicolumn{4}{|c|}{$^{16}$O}\\
  & RH  & RHF & RDDH \\
&&&\\
\hline
&&&\\
$I_L$ & 2.83 & 2.83 & 2.81 \\
$I_T$ & 3.30 & 3.31 & 3.12 \\
$I_{LT}$ & -2.43 & -2.48 & -2.13 \\
$I_{TT}$  & 0.28 & 0.29 & 0.24 \\
&&&\\
\end{tabular}
\caption{The integrated strength $I_i$ of the various response functions
in dimensionless variables. The values calculated according to Eq. 
(\ref{intresponse}) were multiplied by the factor 100.} 
\label{tab:tab2}
\end{table}


\begin{thebibliography}{99}
\bibitem{Frul85} S.~Frullani and J.~Mougey,
                Adv.~Nucl.~Phys.~{\bf 14}, 1 (1985).
\bibitem{Boff93} S. Boffi, C. Giusti and F. D. Pacati,
                 Phys. Rep. {\bf 226}, 1 (1993).
\bibitem{Kell96} J.J. Kelly, Adv.~Nucl.~Phys.~{\bf 23}, 75 (1996).
\bibitem{Mcd90}  J.P. McDermott, Phys. Rev. Lett. {\bf 65} 1991 (1990).
\bibitem{Udia93} J.M. Ud\'\i as, P. Sarriguren, E. Moya de Guerra, E. Garrido
                 and J.A. Caballero, Phys. Rev. C {\bf 48} 2731 (1993).
\bibitem{Jin92}  Y. Jin, D.S. Onley and L.E. Wright, Phys. Rev. C {\bf 45}
                 1311 (1992).
\bibitem{Udia96} J.M. Ud\'\i as, P. Sarriguren, E. Moya de Guerra
                 and J.A. Caballero, Phys. Rev. C {\bf 53} R1488 (1996).
\bibitem{Caba98} J.A. Caballero, T.W. Donelly, E. Moya de Guerra and J.M.
                 Ud\'\i as, Nuc. Phys {\bf A632} 323 (1998).
\bibitem{Sero79} B.D. Serot, Phys. Lett. {\bf 86B}, 146
                 (1979).
\bibitem{Horo81} C.J. Horowitz and B.D. Serot, Nuc. Phys. {\bf A369} 503
                 (1981).
\bibitem{Horo91} C.J. Horowitz, D.P. Murdock and B.D. Serot, in
                 \textit{Computational Nuclear Physics}, edited by
                 K. Langanke, J. A. Maruhn, and S. E. Koonin
                 (Springer-Verlag, Berlin, 1991).
\bibitem{Bouy87} A. Bouyssy, J.-F. Mathiot, Nguyen Van Giai, and S. Marcos,
                 Phys. Rev. C {\bf 36}, 380 (1987).
\bibitem{Fuch95} C. Fuchs, H. Lenske, and H.H. Wolter,
                 Phys. Rev. C {\bf 52}, 3043 (1995).
\bibitem{Frit93} R. Fritz, H. M\"uther, and R. Machleidt,
                 Phys. Rev. Lett {\bf 71}, 46 (1993).
\bibitem{Broc92} R. Brockmann and H. Toki,
                 Phys. Rev. Lett {\bf 68}, 3408 (1992).
\bibitem{Frit94} R. Fritz and H. M\"uther,
                 Phys. Rev. C {\bf 49}, 633 (1994).
\bibitem{Boer94} H.F. Boersma and R. Malfliet,
                 Phys. Rev. C {\bf 49}, 233 (1994); {\bf 50}, 1253(E) (1994).
\bibitem{Don841} T.W. Donelly, in: \textit{Proceedings of
                 the Workshop on Perspectives in Nuclear Physics
                 at intermediate Energies}, edited by S.Boffi,
                 C.C. degli Atti and M. Giannini, World Scientific,
                 Trieste (1984).
\bibitem{Pick85} A. Picklesimer, J.W. Van Orden and S.J. Wallace,
                 Physical Review C {\bf 32}, 1312 (1985).
\bibitem{Don842} T.W. Donelly and A.S. Raskin, Ann. Phys. (N.Y.)
                 {\bf 169}, 247 (1986).
\bibitem{Pick87} A. Picklesimer and J.W. Van Orden,
                 Physical Review C {\bf 35}, 266 (1987).
\bibitem{Boff88} S. Boffi, C. Giusti and F.D. Pacati, 
                 Nuc. Phys. A {\bf 476}, 617 (1988).
\bibitem{Gius89} C. Giusti and F.D. Pacati,
                 Nuc. Phys. A {\bf 476}, 617 (1988).
\bibitem{Rask89} A.S. Raskin and T.W. Donelly, Ann. Phys. (N.Y.)
                 {\bf 191}, 78 (1989).
\bibitem{Pick89} A. Picklesimer and J.W. Van Orden,
                 Physical Review C {\bf 40}, 290 (1989).
\bibitem{Broc90} R. Brockmann and R. Machleidt,
                 Phys. Rev. C {\bf 42}, 1965 (1990).
\bibitem{Hadd93} S. Haddad and M. Weigel,
                 Phys. Rev. C {\bf 48}, 2740 (1993).
\bibitem{deFo83} T. de Forest,
                 Nuc. Phys {\bf A392} 232 (1983).
\end{thebibliography}
\end{document}